\documentclass[structabstract]{aa}
\usepackage{graphicx}
\usepackage{txfonts}
\usepackage{aas_macros}
\usepackage[sort]{natbib}
\usepackage{rotating}
\usepackage{subfig}

\begin{document}

\title{The radio properties of infrared-faint radio sources}

\author{Enno Middelberg\inst{1} \and Ray P. Norris\inst{2} \and
  Christopher A. Hales\inst{3,2} \and Nick Seymour\inst{4} \and
  Melanie Johnston-Hollitt\inst{5} \and Minh T. Huynh\inst{6} \and
  Emil Lenc\inst{2} \and Minnie Y. Mao\inst{7}}

\institute{Astronomisches Institut, Ruhr-Universit\"at Bochum, Universit\"atsstr. 150, 44801 Bochum, Germany\\ \email{middelberg@astro.rub.de} \and
          CSIRO Australia Telescope National Facility, PO Box 76, Epping, NSW, 1710, Australia                                                 \and
          Sydney Institute for Astronomy, The University of Sydney, NSW 2006, Australia                                                        \and
          Mullard Space Science Laboratory, UCL, Holmbury St Mary, Dorking, Surrey, RH5 6NT, UK                                                \and
          School of Chemical and Physical Sciences, Victoria University of Wellington, PO Box 600, Wellington, New Zealand                     \and
          Infrared Processing and Analysis Center, MS220-6, California Institute of Technology, Pasadena CA 91125, USA                         \and
          School of Mathematics and Physics, University of Tasmania, Private Bag 37, Hobart, 7001, Australia                                
}

\date{Received...}

\abstract {Infrared-faint radio sources (IFRS) are objects that have
  flux densities of several mJy at 1.4\,GHz, but that are invisible at
  3.6\,$\mu$m when using sensitive \textit{Spitzer} observations with
  $\mu$Jy sensitivities. Their nature is unclear and difficult to
  investigate since they are only visible in the radio.}
  {High-resolution radio images and comprehensive spectral coverage
    can yield constraints on the emission mechanisms of IFRS and can
    give hints to similarities with known objects.}
  {We imaged a sample of 17 IFRS at 4.8\,GHz and 8.6\,GHz with the
    Australia Telescope Compact Array to determine the structures on
    arcsecond scales. We added radio data from other observing
    projects and from the literature to obtain broad-band radio
    spectra.}
  {We find that the sources in our sample are either resolved out at
    the higher frequencies or are compact at resolutions of a few
    arcsec, which implies that they are smaller than a typical
    galaxy. The spectra of IFRS are remarkably steep, with a median
    spectral index of $-1.4$ and a prominent lack of spectral indices
    larger than $-0.7$. We also find that, given the IR
    non-detections, the ratio of 1.4\,GHz flux density to 3.6\,$\mu$m
    flux density is very high, and this puts them into the same regime
    as high-redshift radio galaxies.}
  {The evidence that IFRS are predominantly high-redshift sources
    driven by active galactic nuclei (AGN) is strong, even though not
    all IFRS may be caused by the same phenomenon. Compared to the
    rare and painstakingly collected high-redshift radio galaxies,
    IFRS appear to be much more abundant, but less luminous,
    AGN-driven galaxies at similar cosmological distances.}

\keywords{Galaxies: active, Galaxies: high-redshift}

\maketitle

\section{Introduction}
Infrared-faint radio sources (IFRS) are radio sources discovered in
deep radio surveys with co-located deep infrared data
(\citealt{Norris2006a}, \citealt{Middelberg2008a},
\citealt{Garn2008}).  A small fraction of the radio sources (about
2\,\%) in these surveys were found to have no identifiable infrared
(IR) counterparts in sensitive observations with the \textit{Spitzer}
infrared telescope, as part of the SWIRE survey ({\it Spitzer}
Wide-Area InfraRed Extragalactic survey,
\citealt{Lonsdale2003}). Given a radio-survey 5\,$\sigma$ sensitivity
of 100\,$\mu$Jy and a 3.6\,$\mu$m (the most sensitive {\it Spitzer}
band) non-detection at a 3-sigma level of 3\,$\mu$Jy, the ratio of
radio flux density to IR flux density, S20/S3.6, is at least 30, so
this S20/S3.6 value is a loose definition of an IFRS. However, most
IFRS have S20/S3.6 values of a few hundred, and some of a few
thousand, when they have radio flux densities of tens of mJy in the
presence of $\mu$Jy-sensitivity IR data. Stacking \textit{Spitzer}
images at the position of the IFRS has failed to show infrared
counterparts, imposing low limits on the IR flux densities and showing
that these are not simply objects that fall just below the
\textit{Spitzer} sensitivity limit (\citealt{Norris2006a},
\citealt{Garn2008}, \citealt{Norris2010}). Since their discovery in
2006, several publications have attempted to understand their nature
and emission mechanisms.\\

There is a growing body of research linking the IFRS phenomenon to
high-redshift active galactic nuclei (AGN). In several publications,
SED modelling of IFRS has been presented, showing that only AGN-driven
objects, redshifted and scaled in luminosity, agree with the
observational evidence.

Very long baseline interferometry (VLBI) observations of a total of
six IFRS by \cite{Norris2007b} and \cite{Middelberg2008c} resulted in
the detection of high-brightness temperature cores in two IFRS,
indicating that they contain AGN. \cite{Middelberg2008c} showed that
the luminosity and morphology of the source was consistent with a
compact steep-spectrum (CSS) source at $z>1$.

\cite{Garn2008} modelled the radio spectra of IFRS and showed that
they are in agreement with scaled-down versions of 3C sources at
$z=2-5$. They found that a variety of template spectra were needed to
reproduce the measurements, indicating that the IFRS are not a
single-source population.

Recently, \cite{Huynh2010} investigated those four IFRS in the Great
Observatories Origins Deep Survey/Chandra Deep Field South
(GOODS/CDFS), identified by \cite{Norris2006a}, which are in a region
for which ultra-deep {\it Spitzer} imaging has recently become
available. For two of the four sources, they were able to identify IR
counterparts at 3.6\,$\mu$m, whereas counterparts were still not
visible for the other two. They used four template spectra to model
the data: the starforming galaxy M\,82; the AGN-dominated,
ultra-luminous infrared galaxy (ULIRG) Mrk\,231; the starburst ULIRG
Arp\,220; and the radio-loud quasar 3C\,273. They found that in the
case of the two IR non-detections, only 3C\,273 was able to reproduce
the measurements, and only when its spectrum was redshifted to $z=1-3$
and obscuration was added to the optical regime. In the case of the
two IR detections, only 3C\,273 was able to fit the data when it was
assumed to be at $z=1.5-2.0$ and an old stellar population was added
to the 3C\,273 spectrum. In no case were the other three template
spectra successful models.

\cite{Norris2010} extended this work using deep {\it Spitzer} imaging
data from the Spitzer Extragalactic Representative Volume Survey
project (\citealt{Lacy2010}), and showed that most IFRS sources have
extreme values of S20/S3.6, which is best fitted by a high-redshift
($z>3$), radio-loud galaxy or quasar. They point out that such AGN are
left as the only viable explanation for the IFRS phenomenon. Local AGN
at moderate ($z<1$) redshifts with host galaxies weak enough to escape
detection by {\it Spitzer} are unknown, and galaxies with AGN and
sufficient dust extinction to obscure the host would require
extinctions in excess of $A_{\rm V}=100^m$ (Arp\,220 is still
IR-bright despite this extinction).

\cite{Huynh2010} and \cite{Norris2010} also consider the possibility
that IFRS are pulsars. While it cannot be ruled out that pulsars are
among the IFRS (in fact, a pulsar is likely to look like an IFRS), the
density of pulsars at high galactic latitudes is of the order of
0.5\,deg$^{-2}$ (\citealt{Manchester2005}), so they cannot account for
the majority of the IFRS.\\

A significant population of AGN which have as yet escaped detection
also has cosmological implications. For example, the cosmic X-ray
background (CXB) has an unresolved component which makes up around
10\,\% in the energy range 0.5\,keV -- 10\,keV
(\citealt{Moretti2003}), and IFRS could in principle account for a
significant fraction of that (\citealt{Zinn2010}). Also models of
structure formation will have to consider a significant additional
fraction of high-redshift AGN.\\

The vast majority of IFRS do not have visible counterparts in
co-located, deep optical images, and no redshift has yet been measured
for an IFRS. Almost the only information available comes from
observations in the radio regime. Useful evidence can be gathered from
broad-band radio spectra, to determine the emission mechanisms at
work. Here we present an analysis of 18 ``bright'' IFRS, using both
archival radio data and new radio observations. The goal of the new
observations was to image a significant sample of sources with high
angular resolution, to determine their structure on kpc scales, and to
obtain more spectral points for an analysis of their spectral indices.

\section{The sample and data}
To construct our sample we used the Australia Telescope Large Area
Survey (ATLAS) 1.4\,GHz catalogues (\citealt{Norris2006a} and
\citealt{Middelberg2008a}). Our targets had been classified as IFRS at
the time of the publication of the catalogues by visual inspection of
the radio and 3.6\,$\mu$m images, and have ratios of 1.4\,GHz flux
density to 3.6\,$\mu$m flux density, S20/S3.6, of between 500 and
10000. Furthermore, existing, yet unpublished 2.4\,GHz survey data
(Zinn et al., in prep) were used to measure the spectral indices,
$\alpha$ ($S\propto\nu^\alpha$), of the sources and to predict their
higher-frequency flux densities. We selected those 18 IFRS with
$S_{\rm 1.4\,GHz}>$1\,mJy which did have reliable 2.4\,GHz detections,
with the exception of CS215, where the 2.4\,GHz emission merged
inseparably with a nearby, strong source. We note that all targets had
a signal-to-noise ratio exceeding 9, so were unquestionably real and
not spurious sources, such as sidelobes. Based on their spectra it was
expected that signal-to-noise ratios of 5 or more could be achieved
with new observations at 4.8\,GHz and 8.6\,GHz for all targets, using
reasonable integration times. Source names are taken from the short
names used by \cite{Norris2006a} and \cite{Middelberg2008a}, with a
prefix of ES for the European Large Area ISO Survey - S1 (ELAIS-S1)
field and CS for the CDFS field.

\subsection{New observations}

The following observations were carried out by us either for this
project only or as part of other observing programmes.

\subsubsection{4.8\,GHz and 8.6\,GHz}
\label{sec:obs}
The targets were observed during five observing runs on 21 to 25
October 2008 when the Australia Telescope Compact Array (ATCA) was in
the 6A configuration. The correlator was configured to allow
simultaneous observations at both 4.8\,GHz and 8.6\,GHz with a
bandwidth of 128\,MHz for each frequency band. In processing, each
band is divided into 13 independent 8\,MHz channels, resulting in an
effective bandwidth of 104\,MHz per band. Three to five sources were
imaged during each observing session, and sources were switched
rapidly to fill the uv plane. More time was spent on weaker targets to
increase the likelihood of detection; the net integration times, after
flagging, are given in Table~1. The flux density scale
was set relative to observations of the primary flux density
calibrator PKS\,B1934-638 with assumed flux densities of 5.828\,Jy at
4.8\,GHz and 2.840\,Jy at 8.6\,GHz. The gain and bandpass calibration
were performed relative to the secondary calibrators PKS\,B0237-233
and PKS\,B0022-423 for the CDFS and ELAIS fields, respectively. Data
calibration was carried out with the Miriad package
(\citealt{Sault1995}) and followed standard procedures as described in
the Miriad User's Guide. Naturally-weighted images with matched
resolution were made at 4.8\,GHz and 8.6\,GHz by excluding the
shortest baseline (CA04-CA05) in the 4.8\,GHz data sets and the
longest three baselines (CA06-CA0[1$|$2$|$3]) at 8.6\,GHz to reduce
the effects of resolution on the measured spectral indices between
these two frequencies. However, resolution effects can still occur
between the lower three frequencies described below and these higher
two frequencies presented here, since the resolutions vary by more
than an order of magnitude. These observations resulted in resolutions
of 4.6$\times$1.7\,arcsec$^2$ on average.

\subsection{2.4\,GHz}
Both the ATLAS/ELAIS and ATLAS/CDFS fields were imaged at 2.4\,GHz
with the ATCA in 2006-2008. The data were imaged and source extraction
and publication is underway (Zinn et al., in prep). For our purpose
here we extracted only the flux densities of the IFRS. The array was
in one of the four available 750\,m configurations due to scheduling
constraints and to ensure that short-spacing information was not
missed at this higher frequency. The observations therefore yielded
much lower resolution than at 1.4\,GHz. The final 1\,$\sigma$ noise
levels are 62\,$\mu$Jy/beam in the ATLAS/ELAIS field and
82\,$\mu$Jy/beam in the ATLAS/CDFS field, owing to a difference in the
integration times. Uniform weighting was used to image both fields,
resulting in resolutions of 33.6$\times$19.9\,arcsec$^2$ and
54.3$\times$20.6\,arcsec$^2$, respectively.

\subsection{Archival data}
The following data were readily available for our sample.

\subsubsection{74 MHz}
The CDFS field is covered by the VLA Low-Frequency Sky Survey (VLSS,
\citealt{Cohen2007}). The VLSS provides images with a resolution of
80\,arcsec and a typical 1\,$\sigma$ noise of 100\,mJy. None of our
sources was detected at 74\,MHz, even though extrapolations from
higher frequencies predicted a marginal detection for at least CS703,
for which a flux density of $S_{\rm 74\,MHz}=490$\,mJy was expected.

\subsubsection{843 MHz}
Flux densities were taken from the Sydney University Molonglo Sky
Survey catalogue (SUMSS, \citealt{Bock1999}) in its 11 March 2008
version. Flux densities are only available for the ELAIS sources since
the SUMSS northern declination limit is $-30^\circ$.
 
\subsubsection{1.4\,GHz}
All sources were observed as part of the ATLAS survey. These
catalogues were made from images with resolutions of the order of
10\,arcsec. To avoid resolution effects when calculating spectral
indices from these data and the 2.4\,GHz data, we convolved the
published, uniformly-weighted images with Gaussian kernels to obtain
the same resolution as at 2.4\,GHz, and re-extracted the flux
densities from the resulting lower-resolution images. Since the
1.4\,GHz observations were carried out with the ATCA in a wide variety
of array configurations, including very compact ones, coverage on
short spacings is excellent.  In some cases this procedure has
slightly increased the flux densities of the sources, which implies
that either these targets were somewhat resolved, or that the
combination of high sensitivity and low resolution was beginning to
show the effects of confusion.

\subsubsection{Flux density errors}

Flux density errors of the 843\,MHz data were taken verbatim from the
catalogue. Other flux densities were modelled as

\begin{equation}
S=g\times(S^\prime + \sigma)
\end{equation}

where $g$ is gain factor which represents the uncertainty of the flux
density scale, derived from the primary calibrator, $S^\prime$ is the
flux density extracted from the image using a Gaussian fit, and
$\sigma$ is the noise in the image. Error propagation then yields a flux
density error of

\begin{equation}
\Delta S=\sqrt{ (g\times\Delta S^\prime)^2 + (g\times \sigma)^2 +  [(S+\sigma)\times\Delta g]^2 }.
\end{equation}
 
$g$ can be set to 1 here since it is only used to express the
uncertainty of the flux density scale, and the expression then reduces
to

\begin{equation}
\Delta S=\sqrt{ \Delta S'^2 + \sigma^2 +  [(S+\sigma)\times\Delta g]^2 }.
\end{equation}

Given the rms of the flux density measurements of our
calibrators\footnote{http://www.narrabri.atnf.csiro.au/calibrators} we
assume that the gain calibration is accurate to $\Delta g$=0.05 and
used $\Delta S'$ as returned by the fitting procedure.

\section{Results and Discussion}

\subsection{The radio properties of IFRS}

\subsubsection{Morphology}

In the case of CS487 the higher-frequency data indicate that the radio
emission is associated with the clearly visible, nearby IR source
SWIRE3\_J033300.99-284716.6, and hence it can no longer be regarded as
an IFRS.  It has therefore been excluded from all analysis presented
in this paper.

Whilst many sources are extended in our 1.4\,GHz images, the
higher-frequency images do not reveal any conclusive substructure, and
since the redshifts of the sources are not known, the conclusions
which can be drawn from this result are limited. If IFRS are
high-redshift sources with z$>$0.5, then one can derive a constraint
on the size of the objects because, in a standard $\Lambda$-dominated
cosmology, the linear size observed in an object varies little beyond
z=0.5: it is 6.08\,kpc/arcsec at z=0.5, rises to 8.56\,kpc/arcsec at
z=1.7 and then drops slowly to 6.41\,kpc/arcsec at z=5. Here we adopt
an average scale of 7\,kpc/arcsec. With restoring beams of
4.6$\times$1.7\,arcsec$^2$ our observations can only resolve
structures larger than about 32\,kpc$\times$12\,kpc, which is only
slightly smaller than a typical galaxy. Some IFRS are very compact in
all images (e.g., CS703, ES427, ES509). In such cases one can conclude
that the sources are smaller than about 1/5 of the highest resolution,
because any extent larger than that would have a noticeable effect on
the image. Therefore these sources must be smaller than
0.9$\times$0.3\,arcsec$^2$, or 4.5\,kpc$\times$2.1\,kpc, which rules
out that these particular sources are simply radio galaxy lobes.
 
We note that classical high redshift radio galaxies have observed
angular sizes that imply projected physical sizes from a few to many
hundreds of kiloparsecs (\citealt{Carilli1994b},
\citealt{Pentericci2000}).  The unresolved sources we see here imply
that they are either intrinsically much smaller than those sources or
that any extended emission has been resolved out, even in the lower
resolution observations.  Given that our sources are an order of
magnitude fainter than the classical radio galaxies, which are 20\,mJy
to 1,000\,mJy, and given that core fractions measured by
\cite{Carilli1994b} and \cite{Pentericci2000} range from $30\%$ to
$<1\%$, IFRS could be analogous to high redshift radio galaxies, but
with radio lobes not bright enough to be seen even in lower resolution
data.

The 1.4\,GHz luminosity of IFRS in our sample, with flux densities
between 1.5\,mJy and 22\,mJy, is in the range of $5\times10^{25}\,{\rm
  W\,Hz^{-1}}$ to $5\times10^{27}\,{\rm W\,Hz^{-1}}$, assuming their
redshifts are between 2 and 5. This predominantly puts them into the
regime of Fanaroff-Riley type II (FR\,II) radio galaxies. We note
however that the break luminosity between FR\,I and FR\,II radio
galaxies depends on the optical luminosity of the host galaxy
(\citealt{Ledlow1996}). At absolute magnitudes of ${\rm M}=-21$, the
break is at $L_{1.4}=10^{24}\,{\rm W\,Hz^{-1}}$, whereas at ${\rm
  M}=-24$ it is two orders of magnitude higher, at
$L_{1.4}=10^{26}\,{\rm W\,Hz^{-1}}$. The IFRS have magnitudes of more
than ${\rm R}=24.5$ (the 95\,\% completeness limit of the co-located
optical observations), hence their absolute magnitudes are greater
than ${\rm M}=-21.5$ at $z=2$ and greater than ${\rm M}=-23.9$ at
$z=2$. At redshifts of 5, all IFRS would exceed a 1.4\,GHz luminosity
of $10^{26}\,{\rm W\,Hz^{-1}}$, so could safely be classified as
FR\,II objects, independent of the optical luminosities of their host
galaxies. At redshifts of 2, however, only the brighter IFRS reach
$10^{26}\,{\rm W\,Hz^{-1}}$, and for those with smaller 1.4\,GHz
luminosities this classification can not be made.

\subsubsection{Spectral indices}
We measured the spectral indices by fitting a power-law to all
available radio data for each source, weighting the data points by
their errors, and ensuring that the data were convolved to the same
beam size as far as this was possible (see Section~\ref{sec:obs}). In
cases where only two data points were available the spectral index was
calculated using these flux densities, and errors were calculated
using error propagation. These values are given in
Table~1.

\begin{figure*}[htpb]
\center
\includegraphics[width=0.45\linewidth]{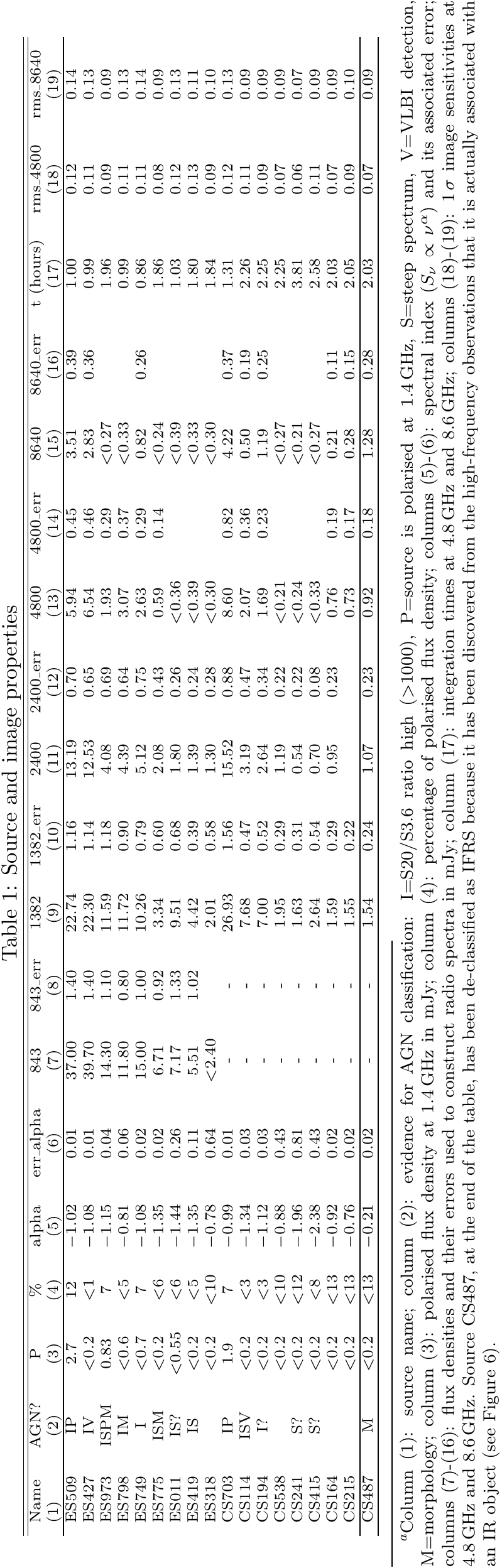}
\end{figure*}

\begin{figure}[htpb]
\includegraphics[width=\linewidth]{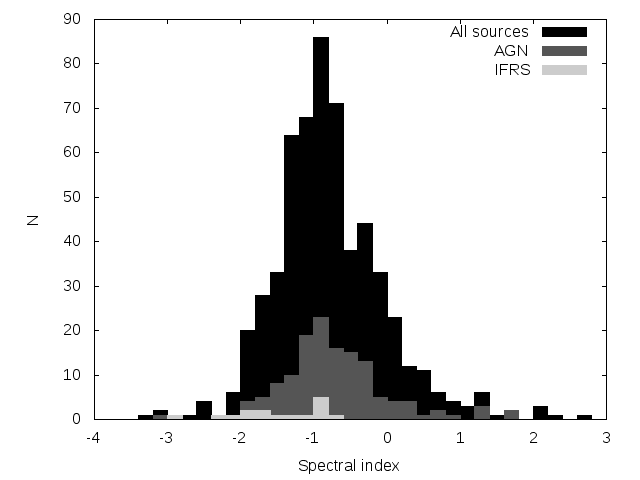}
\caption{Distribution of the 1.4\,GHz to 2.4\,GHz spectral indices
  found in the general source population, in the subsample containing
  AGN, and in our IFRS sample.}
\label{fig:spix}
\end{figure}

To compare the distribution to other sources we calculated the
spectral index using the 1.4\,GHz and 2.4\,GHz data only. A histogram
of the distribution of spectral indices is shown in
Figure~\ref{fig:spix}, along with the spectral indices between
1.4\,GHz and 2.4\,GHz of all sources in the ELAIS field (Zinn et al.,
in prep) and of the AGN contained therein, which were classified based
on morphology, spectroscopy, or radio excess over the radio-IR
relation (see \citealt{Norris2006a} and
\citealt{Middelberg2008a}). The median spectral index of the general
source population in the ELAIS field is $-0.86$, the median of AGN
spectral indices is $-0.82$, and the median of the IFRS is
$-1.40$. The distribution of the IFRS is clearly biased towards low
values, and the tail of indices larger than $-0.7$ is missing
completely. A two-tailed Kolmogorov-Smirnov test shows that the IFRS
distribution differs significantly from the general population
($p=0.0028$) and also from the general AGN population
($p=0.0014$). Since there is plenty of evidence that IFRS are
AGN-driven, the difference in spectral index between the AGN and IFRS
populations must arise from IFRS having rather peculiar properties,
which show up because they have been selected by IR faintness. IFRS
could be AGN in a younger evolutionary stage, at higher redshifts, or
in different environments. We note that the general AGN population
also contains numerous subclasses such as compact steep-spectrum
sources (CSS) and gigahertz-peaked spectrum sources (GPS), which have
peculiar spectral energy distributions, but are not considered
separately in this analysis.

\subsubsection{Changes in the spectral index}

The 4.8\,GHz and 8.6\,GHz observations have higher resolution than the
1.4\,GHz and 2.4\,GHz observations, and are less sensitive to extended
emission. The spectral index between 4.8\,GHz and 2.4\,GHz is
therefore not physically meaningful, because the data at the higher
frequency are sensitive to more compact structures than the 1.4\,GHz
and 2.4\,GHz observations. However, within each pair of bands
(1.4\,GHz/2.4\,GHz or 4.8\,GHz/8.6\,GHz) the uv coverage has been
matched and so the spectral indices are physically relevant to the
size scale being studied.
 
\begin{figure}[htpb]
\includegraphics[width=\linewidth]{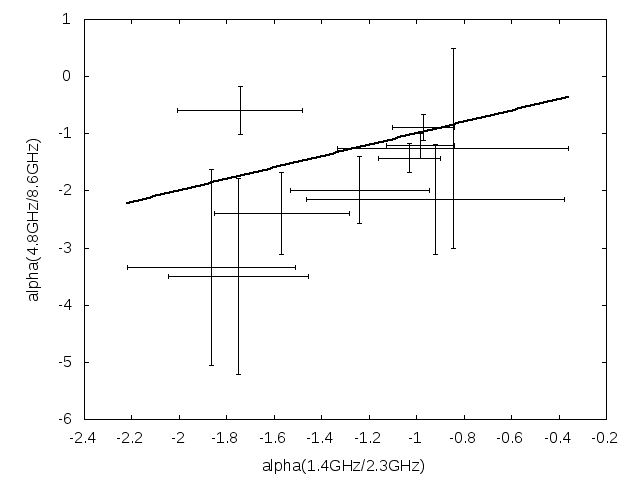}
\caption{The high-frequency spectral index, $\alpha_{4.8}^{8.6}$,
  plotted as a function of the low-frequency spectral index,
  $\alpha_{1.4}^{2.4}$. The solid line indicates where the two would
  be equal (i.e., no change of spectral index with frequency). There
  is a trend that the spectral index steepens towards higher
  frequencies.}
\label{fig:spix_hi_lo}
\end{figure}

We compared the low-frequency (1.4\,GHz/2.4\,GHz) and high-frequency
(4.8\,GHz/8.4\,GHz) spectral indices of 10 targets, 7 of which have
measured flux densities at 1.4\,GHz, 2.4\,GHz, 4.8\,GHz, and 8.6\,GHz,
and 3 of which have upper limits at 8.6\,GHz. In these cases, we used
3 times the image rms as an upper limit on the flux density to compute
the spectral index (the comparatively small span in frequency enlarges
the error bars in these cases). We show in Figure~\ref{fig:spix_hi_lo}
these two spectral indices and indicate with a straight line where
they would be equal. Clearly the spectra steepen towards higher
frequencies.

We note that the median $\alpha_{1.4}^{2.4}=-1.40$ of all IFRS is
lower than the median $\alpha_{1.4}^{2.4}=-1.14$ of the 10 sources
which also have a measurement or limit for $\alpha_{4.8}^{8.6}$. This
is (i) because of the selection effect that the very steep-spectrum
sources tend to have escaped detection at the higher frequencies; and
(ii) because of the use of upper limits at 8.6\,GHz, meaning that the
true spectral index in these three cases is lower than specified by
us.

\subsubsection{Discrepancy between low-frequency spectral index and higher-frequency lower limits}

In some cases (CS538, ES318, ES419, ES749, ES798, and ES973) the
spectral index derived from lower-frequency observations predicts
4.8\,GHz or 8.6\,GHz flux densities which are incompatible with the
measurements at these frequencies. However, in some other cases such
as CS703, ES427, or ES509 the detections at the highest frequencies
align very well with the lower frequencies, and tightly follow
power-laws. We consider this as evidence that the calibration is not
systematically wrong since the same methods were used in all
cases. Instead we consider two effects as potential causes of this
discrepancy.

(i) Sources are resolved out. The 843\,MHz, 1.4\,GHz and 2.4\,GHz data
have excellent uv coverage at spacings below 5\,k$\lambda$ (which
corresponds to an angular scale of 41.25\,arcsec), and even have good
coverage at spacings shorter than 1\,k$\lambda$ (3.43\,arcmin). On the
other hand, in our matched-resolution images at 4.8\,GHz and 8.6\,GHz
the shortest baseline used was 7\,k$\lambda$ (29.5\,arcsec - note that
one goal of these observations was to image the targets with high
resolution, hence long baselines were selected). This means that even
tapered images cannot reveal large-scale structure since this
information simply is not in the data. However, when the angular
resolution is converted to a linear scale one finds that to resolve
objects out in these cases they must be relatively large. At redshifts
of 0.1, 0.5 and 1, a resolution of 29.5\,arcsec corresponds to
54\,kpc, 179\,kpc, and 237\,kpc, respectively (using $H_0=71.0$,
$\Omega_{\rm M}=0.27$, $\Omega_{\rm vac}=0.73$). Only jets and lobes
of radio galaxies are so large, and this would have been noticed in
the 1.4\,GHz images, as is the case in ES011. However, the majority of
the sources suffering from high-frequency drop-outs are not visibly
resolved, and given these size constraints we conclude that resolution
is not the dominant effect. Furthermore, resolution can only explain
spectra where both the 4.8\,GHz and 8.6\,GHz sensitivity limits are
lower than what would be expected from the lower-frequency data.

(ii) In cases where \emph{only} the 8.6\,GHz limit is below the
extrapolation (as in, e.g., ES798 and ES973) resolution effects cannot
account for the observed discrepancy because we carefully selected
baselines such as to obtain matched-resolution images. In these cases,
the spectral steepening must be related to the distribution of the
particle energies in the source. In general, a population of particles
with a power-law distribution of energies (with index $q$) in a
uniform magnetic field will result in synchrotron emission with a
constant radio spectral index with $\alpha=(1-q)/2$. Unless fresh
particles are continuously injected into the source, the particles
with higher energies will lose their energy faster than lower-energy
particles, and the spectrum will steepen. Hence steeper spectra at
higher frequencies could indicate that the AGN has recently been
inactive.

CSS/GPS sometimes show a change in spectral slope at frequencies of
1\,GHz to 10\,GHz (e.g., \citealt{Readhead1996},
\citealt{Murgia2003}), which is commonly attributed to synchrotron
losses (\citealt{Readhead1996}). In our subsample of 10 sources where
independent measurements were available for $\alpha_{1.4}^{2.4}$ and
$\alpha_{4.8}^{8.6}$, the median spectral index changes from $-1.14$
to a median of $-1.71$. This difference of $\Delta\alpha=0.57$ is
comparable to the change in $\alpha$ reported by \cite{Readhead1996}
in the prototypical CSS source COINS\,J2355+4950, which was found to
be 0.6. Whilst this similarity is prone to coincidence and
small-number statistics we note that \cite{Middelberg2008a} already
pointed out the similarities between CSS sources and a VLBI-detected
IFRS.

\subsubsection{Polarisation}
We also searched for polarised emission in our targets. We are in the
process of making a careful analysis of the polarisation levels in the
ATLAS radio data, accounting correctly for the positivity bias in
images of polarised intensity (Hales et al. 2010, in prep). A
preliminary result is that three sources in our sample are
significantly polarised at 1.4\,GHz, whereas upper limits were
obtained for the remaining 14 sources. The polarised sources are ES509
(P=2.70$\pm$0.03\,mJy), ES973 (P=0.83$\pm$0.06\,mJy), and CS703
(P=1.90$\pm$0.05\,mJy). All other sources are unpolarised at the
various levels indicated in Table~1. The limits are
given at $>$99\% credibility (Bayesian confidence) that sources are
unpolarised (\citealt{Vaillancourt2006}).

\subsubsection{Radio-$24\,\mu m$ flux density ratio}

It is well-established that there is a tight correlation, extending
over five orders of magnitude, between the radio and far-infrared
(FIR) luminosity, or flux density, of star-forming galaxies
(\citealt{vanderKruit1973}, \citealt{Condon1982},
\citealt{Dickey1984}, \citealt{deJong1985}). It is attributed to
massive star formation, which generates far-infrared emission by
heating dust, and generates radio emission by accelerating cosmic rays
that then generate radio synchrotron radiation. However, since
far-infrared telescopes are relatively insensitive, the 24\,$\mu$m
flux density is often used as an imperfect proxy for the FIR flux
density \citep{Appleton2004}. While the 24\,$\mu$m flux density is
subject to other emission mechanisms than warm dust, especially at
high redshift, it remains clear (\citealt{Seymour2008}) that a high
value of the radio-24\,$\mu$m flux density ratio indicates the
presence of an AGN

We note that since all our IFRS targets have $S_{\rm 1.4\,GHz}>$1\,mJy
and are undetected not only at 3.6\,$\mu$m but also at 24\,$\mu$m,
with a 5\,$\sigma$ limit of $S_{\rm 24\,\mu m}=252\,\mu$Jy, their
q$_{24}$=log($S_{\rm 24\,\mu m}$/$S_{\rm 20\,cm}$) values are lower
than log(252\,$\mu$Jy/1000\,$\mu$Jy)=$-0.60$. The radio-IR relation for
star forming galaxies has been determined to yield
q$_{24}$=0.84$\pm$0.28 (\citealt{Appleton2004}), so all IFRS have a
more than tenfold radio excess over this relation. The common
interpretation of this is that synchrotron radiation is being produced
without IR emission, which then is regarded as evidence for
non-thermal emission from an AGN. Therefore, all IFRS can be
classified as AGN based on q$_{24}$ alone.

\subsection{IFRS and high-redshift radio galaxies}

There are two main tools for finding radio galaxies at high
redshifts. The so-called z-$\alpha$ relation is derived from the
observation that steep-spectrum radio galaxies tend to have higher
redshifts (\citealt{DeBreuck2002,Klamer2006}). Many high-redshift
radio galaxies (HzRG) have been found exploiting this relation. The
other tool is the K-z relation (\citealt{Lilly1984}) which states that
the logarithm of an object's redshift is proportional to the near-IR
K-band magnitude at 2.2\,$\mu$m. A combination of these two criteria
can be used as an efficient filter for HzRG.

\subsubsection{The ratio of the 1.4\,GHz and 3.6\,$\mu$m flux densities}

We used the widely-used {\it Spitzer} 3.6\,$\mu$m band as a proxy for
K-band observations.  It has been argued previously
(\citealt{Middelberg2008c}, \citealt{Garn2008}, \citealt{Huynh2010},
\citealt{Norris2010}) that the SEDs of IFRS are compatible with those
of high-redshift AGN. We therefore compiled S20/S3.6 values for the
sample of 70 high-redshift radio galaxies (HzRG) by
\cite{Seymour2007}, to compare them to the general radio population
and the IFRS. \cite{Seymour2007} selected from the literature radio
galaxies above a redshift of one with a 3\,GHz luminosity of more than
10$^{26}$\,W/Hz, and supplemented their sample with new or archival
{\it Spitzer} data. IFRS are selected based on the ratio of the radio
and IR flux densities. Sources in the ATLAS radio catalogues typically
have flux densities exceeding 100\,$\mu$Jy (5\,$\sigma$), whereas the
co-spatial 3.6\,$\mu$m observations have 1\,$\sigma$ sensitivities of
around 1\,$\mu$Jy. Therefore a detected radio source with no
catalogued or visibly identifiable counterpart (i.e., $S_{\rm 3.6\,\mu
  m}<3\,\sigma=3\,\mu$Jy) typically has S20/S3.6$>$30. However, the
median S20/S3.6 ratio of the IFRS in our sample is 2330, some two
orders of magnitude larger than this minimum. Hence, while sources
with S20/S3.6$>\approx 50$ could be starbursts or AGN-driven, at
S20/S3.6 exceeding a few hundred are likely to be similar to the HzRG.

The median S20/S3.6 ratio of all sources in the ATLAS/ELAIS field
(with detections in both the radio and IR bands) is 6.12, but the
distribution extends over five orders of magnitude. With ratios
between approximately 500 and 10000, the IFRS clearly are at and
beyond the high tail of the distribution of the general source
population.

\begin{figure}[htpb]
\includegraphics[width=\linewidth]{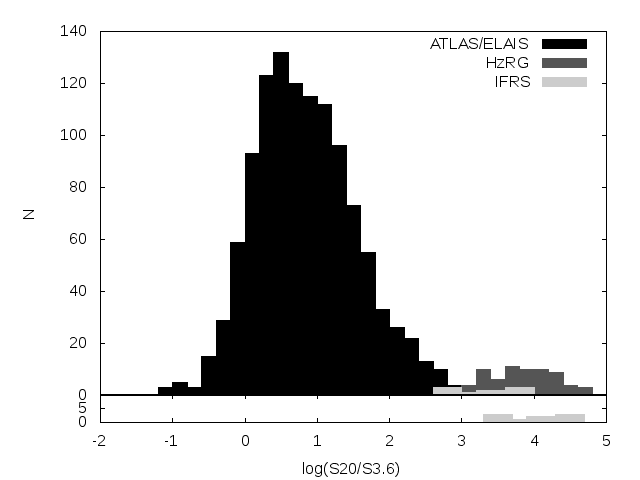}
\caption{{\it Top panel:} The distribution of S20/S3.6 in the general
  source population, in the sample of HzRG by \cite{Seymour2007}, and
  in our IFRS sample. The IFRS clearly occupy a different regime than
  the general population, and tend to overlap more with the HzRG. {\it
    Bottom panel:} The histogram of the IFRS S20/S3.6 ratios as in the
  upper panel, shifted to the right by log(5)=0.7. This takes into
  account that \cite{Norris2010} found no IR counterparts for IFRS in
  a stacking analysis with a 5 times higher sensitivity. On average
  the IFRS then have a S20/S3.6 which is about two times higher than
  that of the HzRG.}
\label{fig:radio-ir}
\end{figure}

The median S20/S3.6 of the HzRG by \cite{Seymour2007} is 6550,
significantly larger than in the general source population, and closer
to the IFRS median. However, HzRG are much brighter -- they present
high luminosities and have been gathered from surveys covering much
larger areas. In contrast, IFRS are fainter (by a factor of around 50
if they are assumed to be at the same redshifts as HzRG) and have a
surface density of around 10\,deg$^{-2}$, whereas that of HzRG is
0.001\,deg$^{-2}$ -- approximately four orders of magnitude smaller. A
histogram of S20/S3.6 of the general radio source population, the HzRG
and the IFRS is shown in Figure~\ref{fig:radio-ir}.

We stress that since for IFRS S20/S3.6 has been calculated using upper
limits on the 3.6\,$\mu$m flux density, the true ratio is expected to
be larger. Given the stacking experiments by \cite{Norris2010}, who
find that the median flux density of IFRS 3.6\,$\mu$m counterparts is
less than 0.2\,$\mu$Jy, we point out that the S20/S3.6 ratios of IFRS
could be as much as a factor of 5 higher than estimated here. This
would shift the IFRS in Figure~\ref{fig:radio-ir} to the right by
log(5)=0.7 (illustrated in the lower panel of
Figure~\ref{fig:radio-ir}). The median S20/S3.6 of the IFRS then
increases to 11650, almost two times the HzRG median.

\cite{Norris2010} also show that IFRS are consistent with radio-loud
AGN at redshifts greater than 3. They used observations from the {\it
  Spitzer} Warm Mission (5\,$\sigma$ limits of 1\,$\mu$Jy, 5 times
deeper than the {\it Spitzer} SWIRE data) and the published ATLAS
radio catalogues. All of their IFRS remain undetected even with these
new observations. They argue that, since the S20/S3.6 ratios are of
the same order of magnitude as the HzRG by \cite{Seymour2007}, and
since the 3.6\,$\mu$m flux densities of HzRG drop below the {\it
  Spitzer} detection limit when at redshifts larger than 3, IFRS are
likely to be at similarly high redshifts.

\cite{Huynh2010} analysed four IFRS in the GOODS/CDFS, which are
located in a region for which very deep {\it Spitzer} data had
recently become available. They find counterparts for only two of the
IFRS, CS446, and CS506, yielding S20/S3.6 ratios of 51.2 and 30.9,
respectively. The two IFRS still undetected with the new {\it Spitzer}
data, CS283 and CS415, have S20/S3.6 ratios of $>$137 and $>$2520,
respectively. CS415 has a radio spectral index of $-1.1$ but was not
included in our sample because it was deemed too faint for successful
4.8\,GHz and 8.6\,GHz observations.

From the histogram of S20/S3.6 it becomes clear that there is more
overlap between IFRS and HzRG than between IFRS and the general radio
source population. IFRS appear to have steep radio spectra and very
faint IR flux densities, both of which indicate that they are
high-redshift AGN.

One could argue that IFRS-like sources exist in the data which are not
classified as such because they are above the SWIRE 3.6\,$\mu$m
detection limit, but have similar S20/S3.6 ratios. There are 16
non-IFRS in the ATLAS/ELAIS catalogue with S20/S3.6 exceeding 500, the
minimum found among the IFRS. All these sources were classified as
AGN, based on either radio morphology or because of their clear
(10-fold) excess over the radio-infrared relation (some do not have a
detected 24\,$\mu$m counterpart, so the 5\,$\sigma$ detection limit of
252\,$\mu$Jy was assumed). There are 66 more sources with S20/S3.6
between 100 and 500, all of which were classified as AGN based on
q$_{24}$=log($S_{\rm 24\,\mu m}$/$S_{\rm 20\,cm}$)
(\citealt{Appleton2004}, using a 252\,$\mu$Jy limit in case of
non-detections). Hence we argue that an excess of radio over
3.6\,$\mu$m, like q$_{24}$, can be regarded as an indicator of
non-thermal emission from AGN, and that this finding supports the
hypothesis that IFRS host AGN. Out of the 82 sources in the
ATLAS/ELAIS catalogue with measured S20/S3.6 in excess of 100, 6 have
a spectroscopic redshift. The median of these redshifts is 0.38, and
only a single object has a redshift larger than 1, with $z=1.82$. We
conclude that objects with measured S20/S3.6 in the same range as IFRS
could indeed be similar objects at lower redshifts.

We also computed S20/S3.6 for the submm-detected sources of the
SCUBA-SHADES survey (\citealt{Clements2008,Ivison2007}). We find that
these sources in general have rather low S20/S3.6 values, of the order
of a few, with only three exceeding 10. This result is consistent with
them being starburst galaxies.

\subsubsection{The spectral indices of IFRS and HzRG}

\begin{figure}[htpb]
\includegraphics[width=\linewidth]{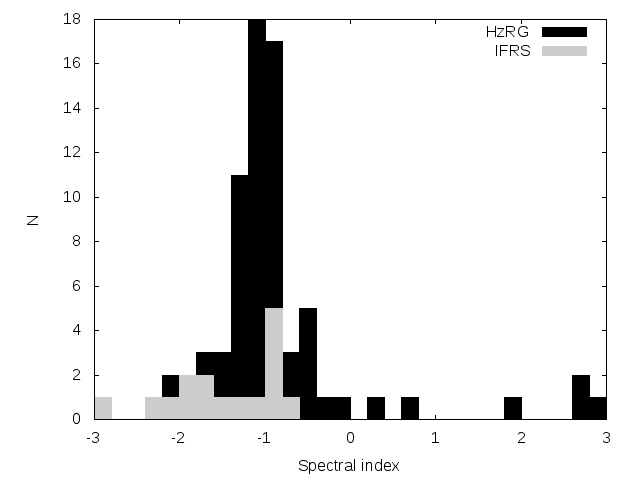}
\caption{The distribution of the spectral indices of IFRS and
  HzRG. The median of the IFRS sample is $-1.4$, and the median of the
  HzRG sample is $-1.02$.}
\label{fig:alphas}
\end{figure}

We show in Figure~\ref{fig:alphas} the distribution of the spectral
indices of IFRS and the spectral indices of the HzRG by
\cite{Seymour2007}. The median of the IFRS sample is $-1.4$, whereas
the median of the HzRG sample is $-1.02$. Given the $z-\alpha$
relation this indicates that IFRS are located at even higher redshifts
than the HzRG. A K-S-test shows with $p=0.011$ that the two
distributions are unlikely to be drawn from the same parent
distribution.

\subsubsection{Simulated surface densities of radio emitters}

The extragalactic part of the SKA simulated skies, S$^3$-SEX
(\citealt{Wilman2010}), is a simulation of radio continuum sources in
an area of $20\times20$\,deg$^2$ out to redshift $z=20$. It represents
today's knowledge about the radio sky and is used as reference when
predictions are needed for the findings with new telescopes such as
Lofar and the SKA pathfinders.

We queried S$^3$-SEX for star-forming galaxies, quasars, and FR\,I and
FR\,II radio galaxies with a 1.4\,GHz flux density exceeding
1\,mJy. In a first query, we searched for sources at $z>2$ and in a
second query for sources with $z>4$. Our findings are listed in
Table~\ref{tab:s-cubed}.

We find that the surface densities of star-forming galaxies and radio
quasars is too low by orders of magnitude to account for the IFRS
phenomenon. FR\,I/II sources, however, have much higher surface
densities, even higher than the number of IFRS in our
sample. Therefore they could account for at least a fraction of the
IFRS. However, the observations presented here and by
\cite{Middelberg2008c} indicate that in general such objects are too
large.

\addtocounter{table}{1}
\begin{table}
\caption{Surface densities of various types of radio sources taken
  from the extragalactic part of the SKA simulated sky. Shown are the
  results for $z>2$ and $z>4$, grouped into star-forming galaxies
  (SF), radio quasars (RQ) and Fanaroff-Riley type I and II (FR\,I/II)
  sources.}
\label{tab:s-cubed}
\centering
\begin{tabular}{lcc}
\hline
\hline
Type & $N_{2<z<4}$ & $N_{z>4}$\\
     & deg$^{-2}$  & deg$^{-2}$\\
\hline
SF     &  0.058 & 0.000 \\
RQ     &  0.033 & 0.000 \\
FR\,I  & 12.162 & 1.207 \\
FR\,II &  3.485 & 0.505 \\
\hline
\end{tabular}
\end{table}

\subsection{Notes on individual sources and their classification}

Here we describe the individual targets. 

\begin{itemize}
 
   \item ES509 is a strong, compact radio source at 1.4\,GHz and is
     detected at all other frequencies. Its spectrum is a well-defined
     power-law with an index of $-1.02$, and its S20/S3.6 ratio is 9240,
     clearly in the realm of the HzRG sample by \cite{Seymour2007}. It
     is significantly polarised at 1.4\,GHz (12\%) and also at higher
     frequencies. We conclude from this evidence that this is an
     AGN. However, this source was not detected by
     \cite{Middelberg2008c} in VLBI observations, who determined that
     its compact flux density was lower than 0.27\,mJy at 1.4\,GHz.

    \item ES427 is similar to ES509 in that it is a strong, compact
      radio source at 1.4\,GHz with a spectral index of $-1.08$, and has
      a S20/S3.6 ratio of 9070. However, there are two significant
      differences to ES509: ES427 is unpolarised in all our radio
      images ($<$1\,\%), but was detected and imaged by
      \cite{Middelberg2008c} using VLBI, who suggested that its
      properties are consistent with it being a CSS. We therefore
      conclude that this is an AGN.

    \item ES973 is an extended radio source at 1.4\,GHz which has been
      classified as AGN based on this finding alone, but its S20/S3.6
      ratio of 4710 supports this. Its spectrum is steep with an index
      of $-1.15$, but it is not detected at 8.6\,GHz, where, according
      to an extrapolation from the lower frequencies, a 13.3\,$\sigma$
      detection would have been expected. It is polarised at 1.4\,GHz
      (7\,\%) and unpolarised at 4.8\,GHz. Therefore, based on
      morphology and polarisation, this is an AGN.

    \item ES798 is a somewhat extended radio source at 1.4\,GHz, and
      the detection at 4.8\,GHz pinpoints a position which suggests
      that the nearby IR source is not associated, as had already been
      suggested by \cite{Middelberg2008a}. Its S20/S3.6 ratio is 4760
      which is high, and its spectral index of $-0.81$ is not
      well-defined. The non-detection at 8.6\,GHz is odd since a
      14.6\,$\sigma$ detection would have been expected. It is
      unpolarised at 1.4\,GHz ($<$5\,\%), but still we deem the
      evidence based on radio morphology and S20/S3.6 sufficient to
      classify this source as an AGN.

    \item ES749 is a somewhat extended radio source at 1.4\,GHz with a
      high S20/S3.6 of 4170. Its extended morphology suggests that it
      is AGN-driven, and this is confirmed by its rather steep
      spectrum with index $-1.08$. Its 8.6\,GHz flux density is lower
      than expected from lower-frequency extrapolation, and it is
      unpolarised ($<$7\%). We classify this source as an AGN.

    \item ES775 is another extended (at 1.4\,GHz) radio source with a
      S20/S3.6 ratio of 1360 and a steep spectrum with index
      $-1.35$. A faint bridge of emission extends towards the nearby
      source ES780, and the nature of this bridge is unclear. ES775 is
      not detected at 8.6\,GHz due to sensitivity limits, and is
      unpolarised ($<$6\,\%). \cite{Middelberg2008c} found that in
      VLBI observations the compact emission from this source was less
      than 0.26\,mJy at 1.6\,GHz, however, the arcsec-scale morphology
      and steep spectrum lead us to conclude that this source is
      AGN-driven.

    \item ES011 is a significantly extended source at 1.4\,GHz with a
      high S20/S3.6 ratio of 3870 and a steep, albeit somewhat
      ill-defined, radio spectrum with index $-1.44$. The 1.4\,GHz
      emission appears too high, part of which we ascribe to confusion
      with nearby sources (the flux density being measured from a
      low-resolution image). If confusion is indeed an issue, the flux
      density at the lower three frequencies could be over-estimated,
      and therefore the non-detections at 4.8\,GHz and 8.6\,GHz would
      not be significant. ES011 is unpolarised ($<$6\,\%), and
      therefore we find no strong evidence that this source contains
      an AGN.
 
   \item ES419 is a compact 1.4\,GHz radio source with a moderately
     high S20/S3.6 ratio of 1800. This source was deemed by
     \cite{Middelberg2008a} not to be associated with the nearby IR
     objects visible in Figure~\ref{fig:images}. Its lower-frequency
     radio spectral index is a rather steep $-1.35$, and it has not been
     detected at the higher frequencies because of its faintness. It
     is unpolarised ($<$5\,\%), and so the only clues to its
     classification are S20/S3.6 and $\alpha$, and we tentatively
     identify this source as an AGN.

    \item ES318 is a slightly extended, faint 1.4\,GHz radio source
      with an ill-defined spectral index of $-0.78$ and a moderate
      S20/S3.6 ratio of 820. It is unpolarised ($<$10\,\%), but we do
      not have sufficient evidence to make a reasonable
      classification.

    \item CS703 is a very compact, strong radio source with a
      well-defined, rather steep spectrum with index $-0.99$ and a high
      S20/S3.6 ratio of 8980. The non-detection at 74\,MHz can
      probably be attributed to synchrotron self-absorption. It is
      polarised at all frequencies (7\,\% at 1.4\,GHz), and since we
      do not see any evidence of it being resolved, we classify this
      source as an AGN.
 
   \item CS114 is a compact radio source at all frequencies with a
     steep spectrum ($\alpha=-1.34$) and a high S20/S3.6 of 2560. It
     is known to be an AGN because of its VLBI-detected compact core
     (\citealt{Norris2007b}) which contains 5\,mJy and thus most of
     the total emission. It is unpolarised ($<$3\,\%), but we suggest
     it is AGN-driven.
 
   \item CS194 is a compact radio source at all frequencies and does
     have a steep spectrum with index $-1.12$ and an S20/S3.6 ratio of
     2330. \cite{Norris2007b} do not detect this source in a VLBI
     observation and conclude that if it contains an AGN its compact
     emission must be less than 1\,mJy. Hence its emission is compact
     enough to be unresolved at a few arcsec, but extended enough to
     be resolved out on the shortest LBA baseline with a resolution of
     390\,mas. It is one of the two sources which have a higher
     spectral index at the higher frequencies, and was not found to be
     polarised ($<$3\,\% at 1.4\,GHz) at any frequency. We do not have
     sufficient evidence to conclusively classify this source as an
     AGN.

    \item CS538 is a faint, slightly extended 1.4\,GHz radio source,
      for which we only have two flux density measurements, yielding a
      spectral index of $-0.88$. Its S20/S3.6 ratio is 650, and the
      non-detections at 4.8\,GHz and 8.6\,GHz are due to its
      faintness. The 1.4\,GHz emission is extended towards a nearby IR
      source, but unless higher-resolution imaging is provided an
      association remains unclear. It is unpolarised at 1.4\,GHz
      ($<$10\,\%). Hence it can not be classified reliably as an AGN.

    \item CS241 is another faint, unpolarised ($<$12\,\%) 1.4\,GHz
      radio source with a remarkably steep, but ill-defined, spectral
      index of $-1.96$ and an S20/S3.6 ratio of 540. Its classification
      remains unclear due to its faintness.

    \item CS415 is a radio source with a remarkably steep spectrum
      ($\alpha=-2.38$), even though this is not well-defined. Its
      S20/S3.6 ratio is still high (880) compared to the general radio
      population, and it is unpolarised ($<$8\,\% at 1.4\,GHz). Its
      classification is unclear.

    \item CS164 is a 1.6\,mJy radio source at 1.4\,GHz with a
      well-defined spectral index of $-0.92$ and an S20/S3.6 ratio of
      530, placing it at the lower end of the IFRS distribution. Like
      most sources it is unpolarised at 1.4\,GHz ($<$13\,\%), and we
      therefore do not have reliable evidence that this is an AGN.

    \item CS215 is a faint radio source with a ``normal'' spectral index
      of $-0.76$ and has at 520 the lowest S20/S3.6 ratio in our
      sample. At 2.4\,GHz the source merges with other nearby radios
      sources and we were unable to measure its flux density
      there. The SED suggests that the 1.4\,GHz data point is
      flattening the spectrum somewhat, but this is speculative. The
      source is unpolarised ($<$13\,\% at 1.4\,GHz), and our evidence
      therefore is not sufficient to classify this as an AGN.
\end{itemize}
 
\section{Conclusions}
We compiled radio observations of a sample of IFRS, and added
dedicated high-frequency, high-resolution observations to investigate
the nature of these objects. Using the IR detection limits we computed
their ratio of 1.4\,GHz to 3.6\,$\mu$m flux density and compared them
with the general radio source population and a sample of high-redshift
radio galaxies. Our conclusions are as follows.
 
\begin{itemize}

    \item All IFRS can be classified as AGN based on their radio
      excess over the radio-IR relation. This is encapsulated in all
      of them having a value of q$_{24}$ of less than $-0.60$.
  
    \item The distribution of the IFRS spectral indices is
      significantly different from the distribution of the general
      radio source population, and also different from the AGN
      population. The spectra are steep, with a median of
      $\alpha=-1.4$, which is much steeper than the spectral indices
      of a sample of HzRG by \cite{Seymour2007}, who found a median of
      $-1.02$.  Also there is a prominent lack of spectral indices
      larger than $-0.7$. The very steep radio spectra indicate a
      fundamental difference to the general radio source population.

    \item We find tentative evidence that radio spectra are steepening
      towards higher frequencies, indicating synchrotron losses. The
      spectra display curvature seen in CSS/GPS sources rather than
      the power-law more typical of classical AGN.

    \item The ratio of 1.4\,GHz flux density to 3.6\,$\mu$m flux
      density, S20/S3.6, is biased to much higher values than that of
      the general radio source population, and has significant overlap
      with the HzRG distribution of the \cite{Seymour2007}
      sample. Whilst the HzRG are very rare sources selected from
      all-sky surveys with very high luminosities, IFRS could be a
      less luminous, but much more abundant version of AGN-driven
      sources at very high redshifts.

    \item Out of the 18 sources discussed here, one (CS487) has been
      de-classified as IFRS because its radio emission showed a solid
      association with a nearby IR source. Out of the 17 remaining
      sources, 10 have been classified as AGN based on either their
      S20/S3.6 ratio (9), polarisation properties (4), radio spectral
      index (4), a VLBI detection (2) and radio morphology (3).

    \item 7 sources have insufficient data to yield a positive AGN
      classification (except for their value of q$_{24}$), mostly
      because our radio data were not sufficiently sensitive.

\end{itemize}

We have presented further evidence that IFRS are a distinct group of
radio sources which are principally detected and studied via radio
observations. Although the brighter IFRS are likely to be similar to
HzRG (\citealt{Seymour2007}, \citealt{Jarvis2001}), they are, on
average, probably much less luminous. So the IFRS probably represent
obscured, radio-loud AGN which have not previously been studied. It is
expected that future radio surveys of the sensitivity of the ATLAS
survey, such as ASKAP-EMU, combined with infrared observations, will
uncover IFRS in the thousands, allowing further examination of their
characteristics.

\begin{figure*}
\includegraphics[width=\linewidth]{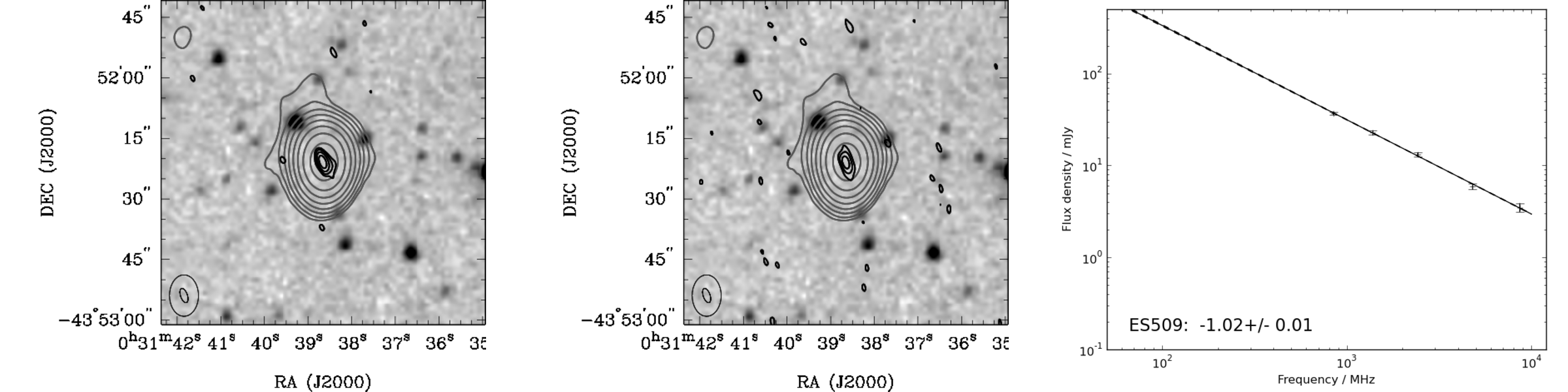}
\includegraphics[width=\linewidth]{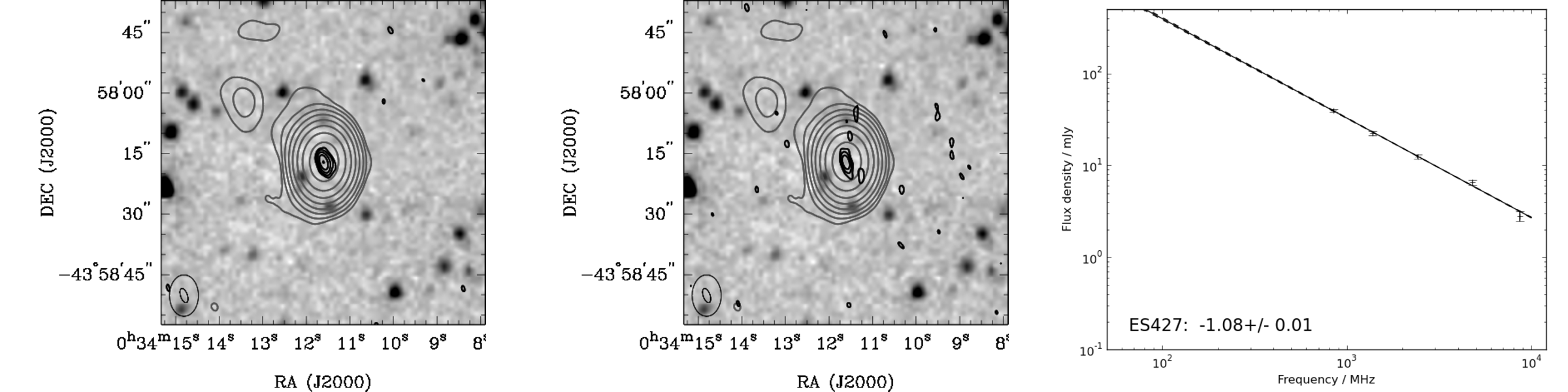}
\includegraphics[width=\linewidth]{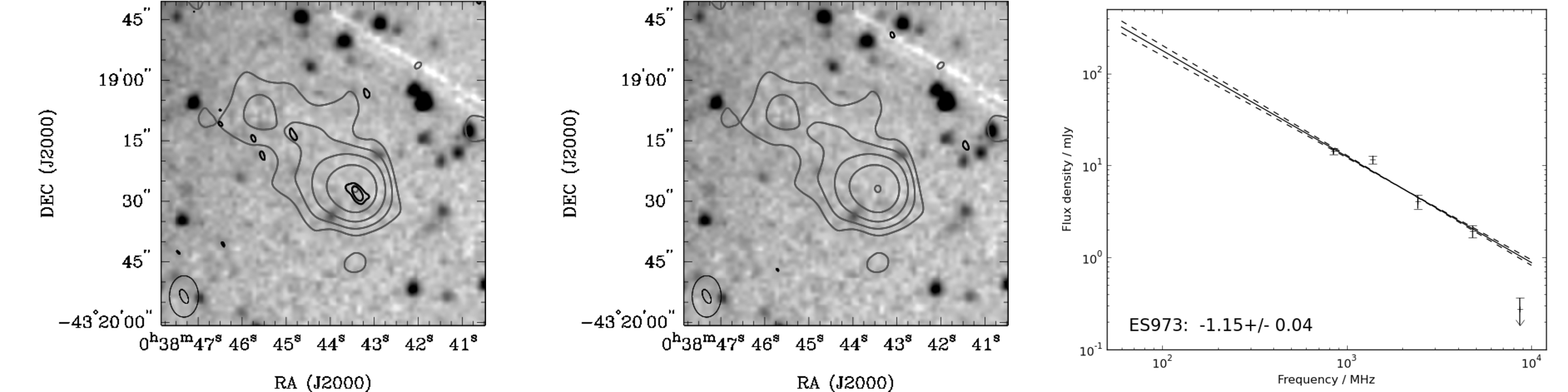}
\includegraphics[width=\linewidth]{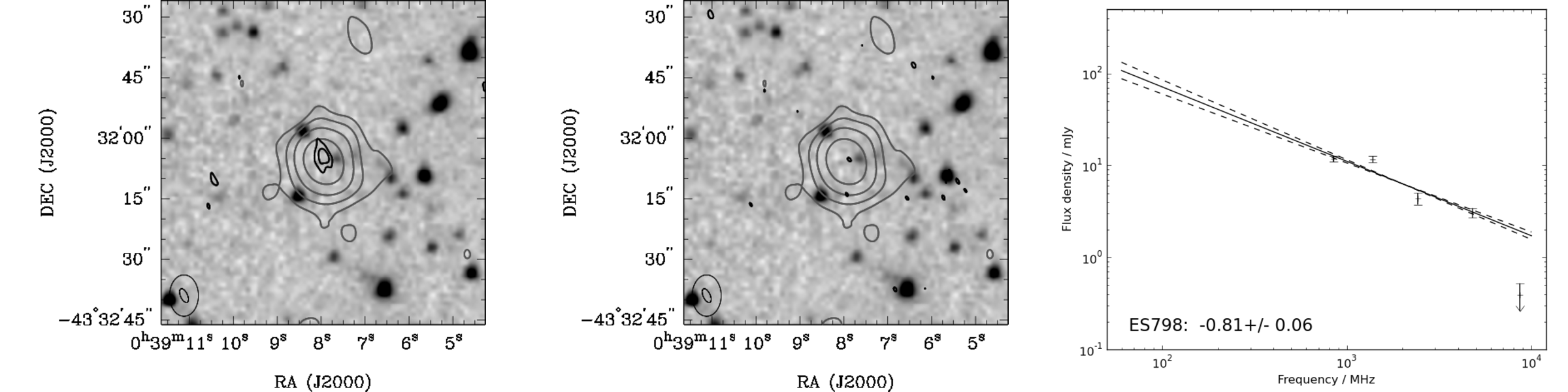}
\includegraphics[width=\linewidth]{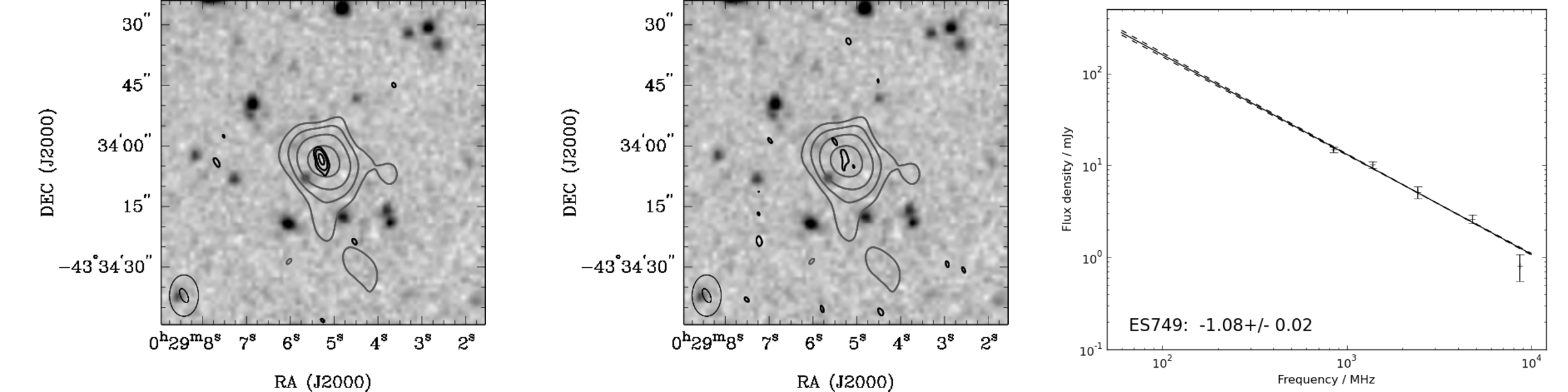}
\caption{Each row shows a grey-scale image of the {\it Spitzer}
  3.6\,$\mu$m observations, superimposed with grey contours indicating
  the 1.4\,GHz observations and black contours showing the 4.8\,GHz
  (left panel) and 8.6\,GHz (middle panel) observations. The IFRS are
  always the sources at the image centres. Contours start at
  3\,$\sigma$ and increase by factors of 2. The 1.4\,GHz restoring
  beam and the 4.8\,GHz/8.6\,GHz restoring beams, which are the same
  size, are indicated with ellipses in the lower left corners of the
  images. The right panel shows the flux density measurements
  available for a source and 3\,$\sigma$ upper limits where no
  detection was made (indicated with arrows). The solid line indicates
  the best available spectral index, and dashed lines indicate a
  power-law with an index 1\,$\sigma$ larger and 1\,$\sigma$ smaller
  than determined by the data. We note that all sources have a
  signal-to-noise ratio of more than 9 in the 1.4\,GHz observations,
  so there is no doubt that they are real sources and not spurious.}
\end{figure*}

\begin{figure*}
\ContinuedFloat
\includegraphics[width=\linewidth]{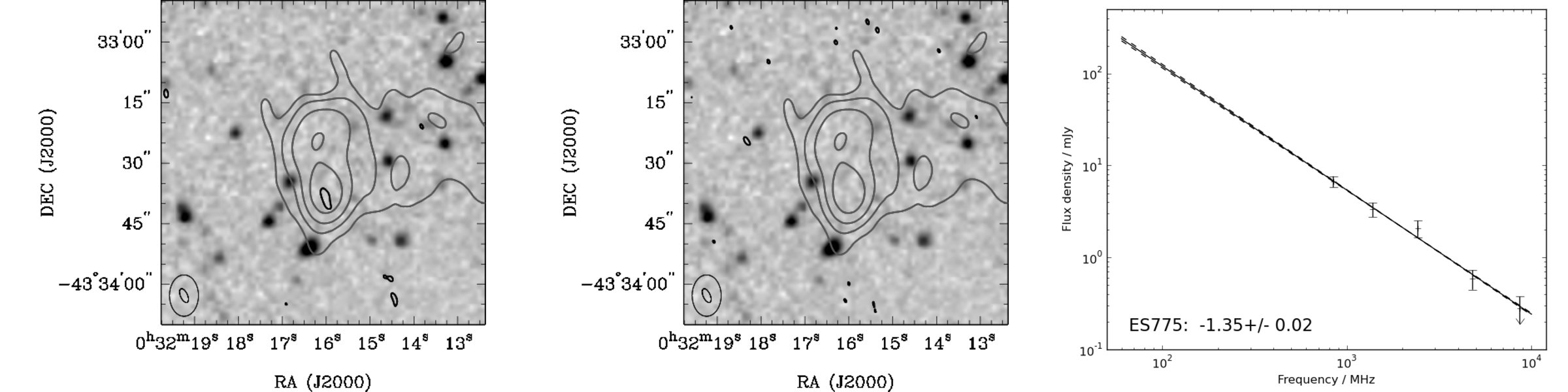}
\includegraphics[width=\linewidth]{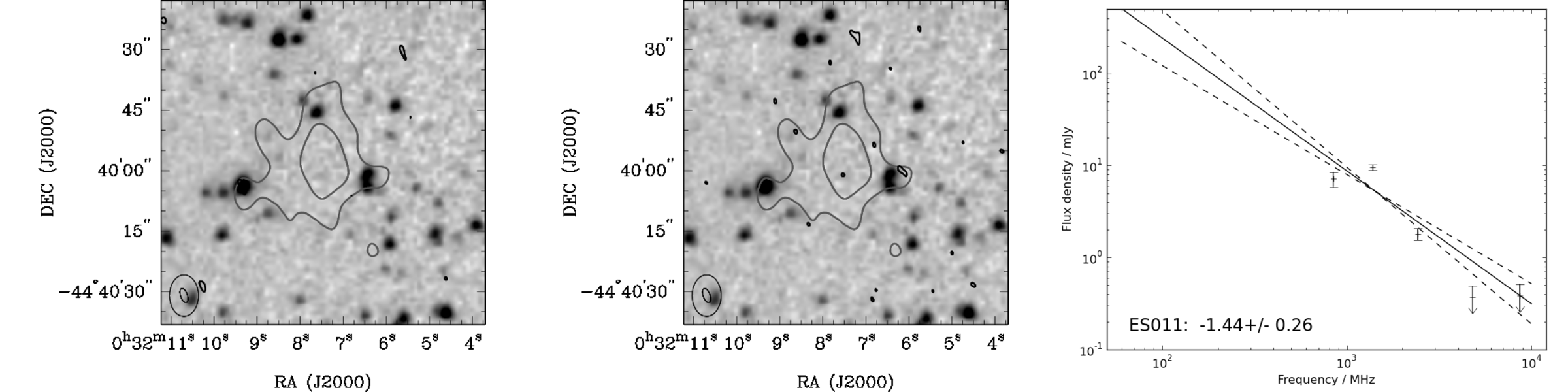}
\includegraphics[width=\linewidth]{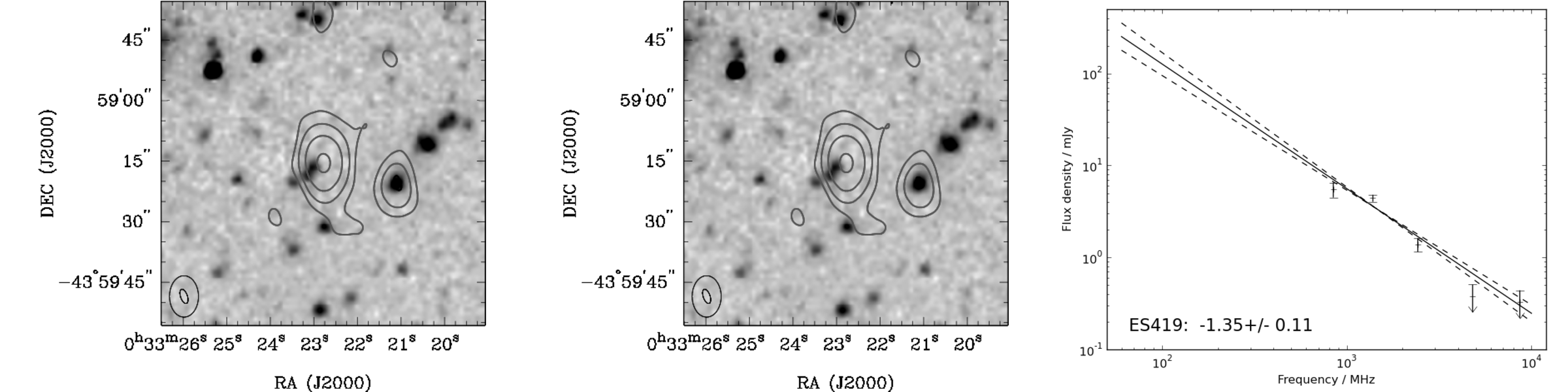}
\includegraphics[width=\linewidth]{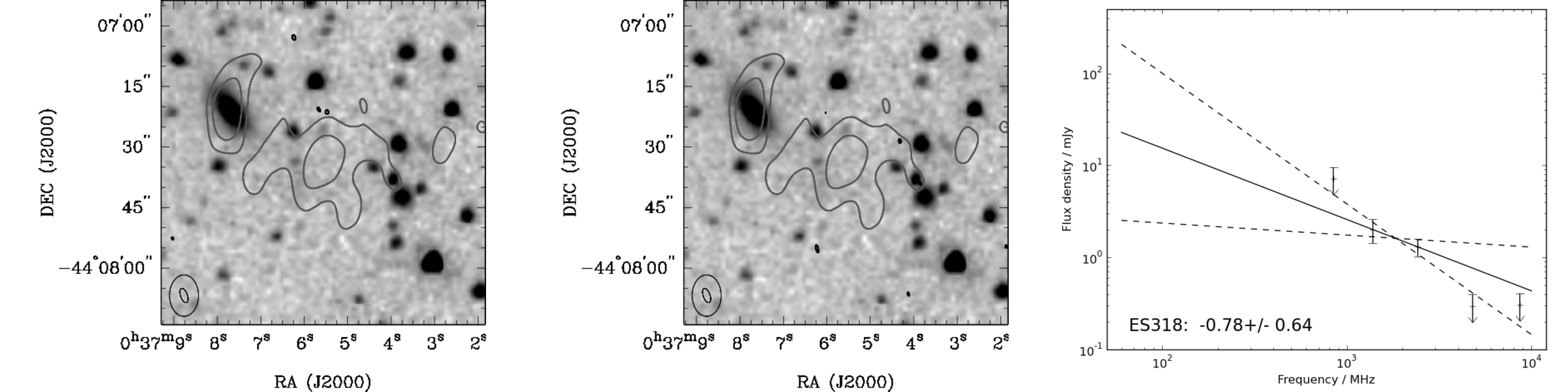}
\includegraphics[width=\linewidth]{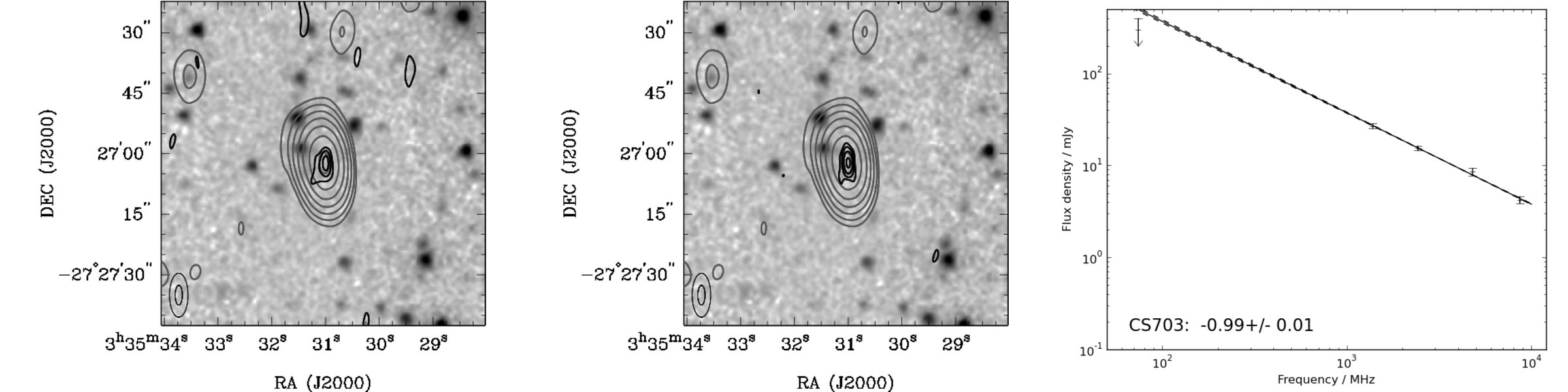}
\caption{(continued)}
\label{fig:images}
\end{figure*}

\begin{figure*}
\ContinuedFloat
\includegraphics[width=\linewidth]{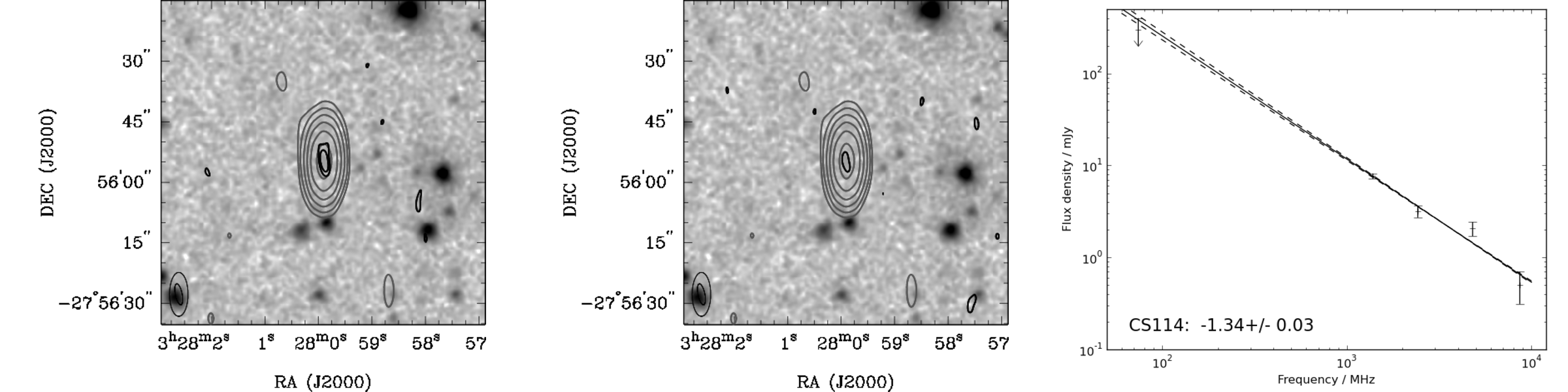}
\includegraphics[width=\linewidth]{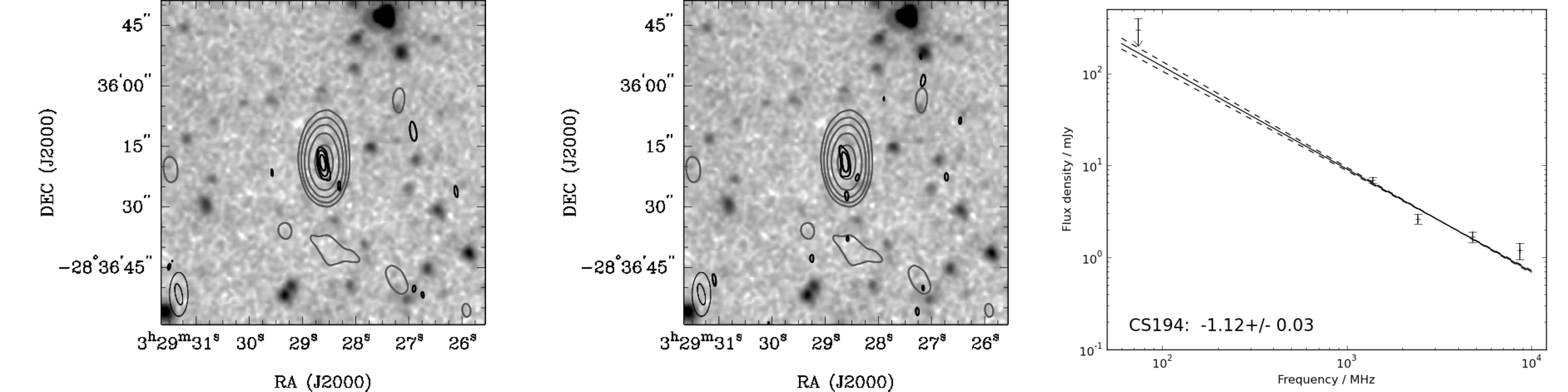}
\includegraphics[width=\linewidth]{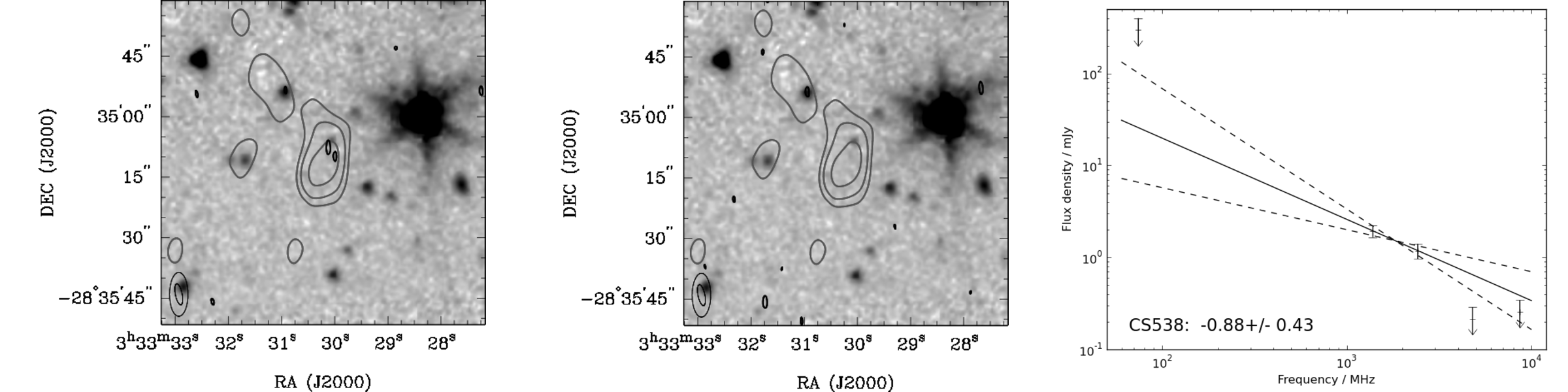}
\includegraphics[width=\linewidth]{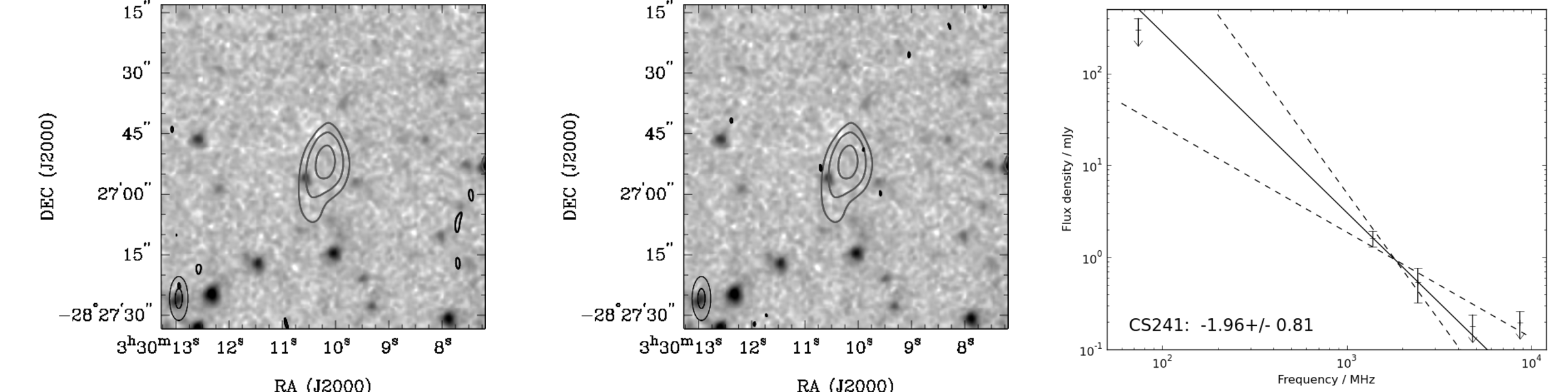}
\includegraphics[width=\linewidth]{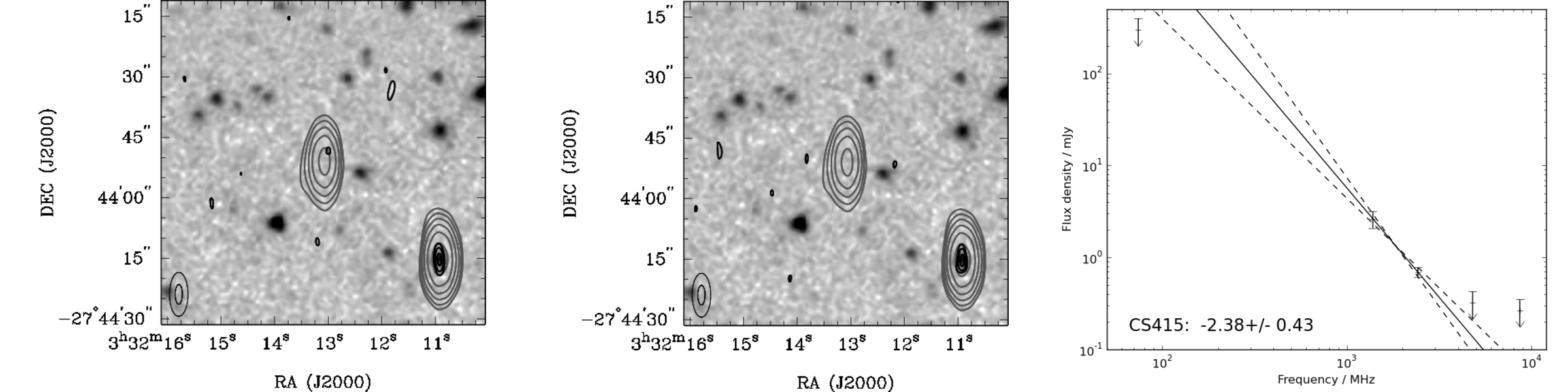}
\caption{(continued)}
\end{figure*}

\begin{figure*}
\ContinuedFloat
\includegraphics[width=\linewidth]{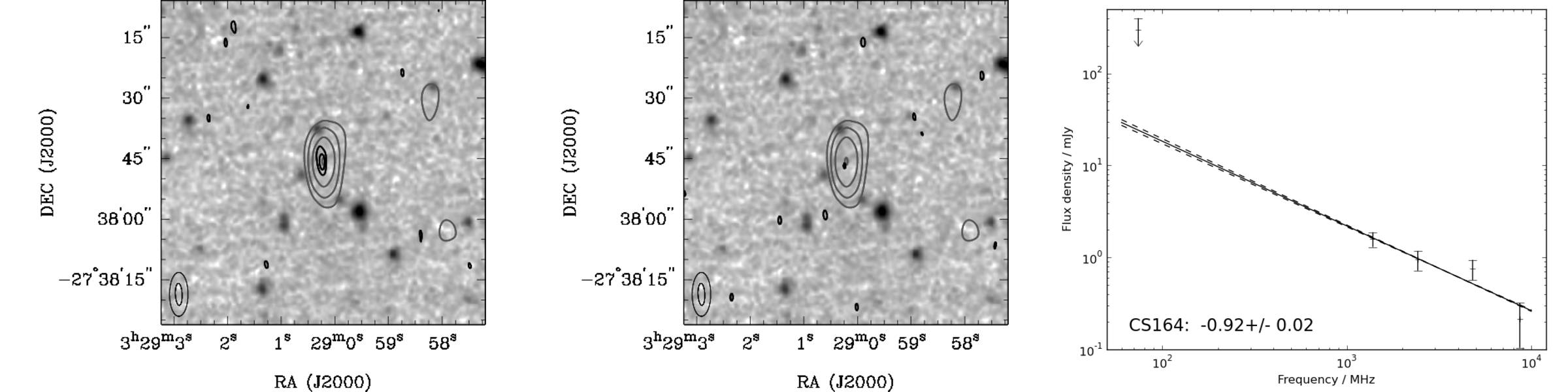}
\includegraphics[width=\linewidth]{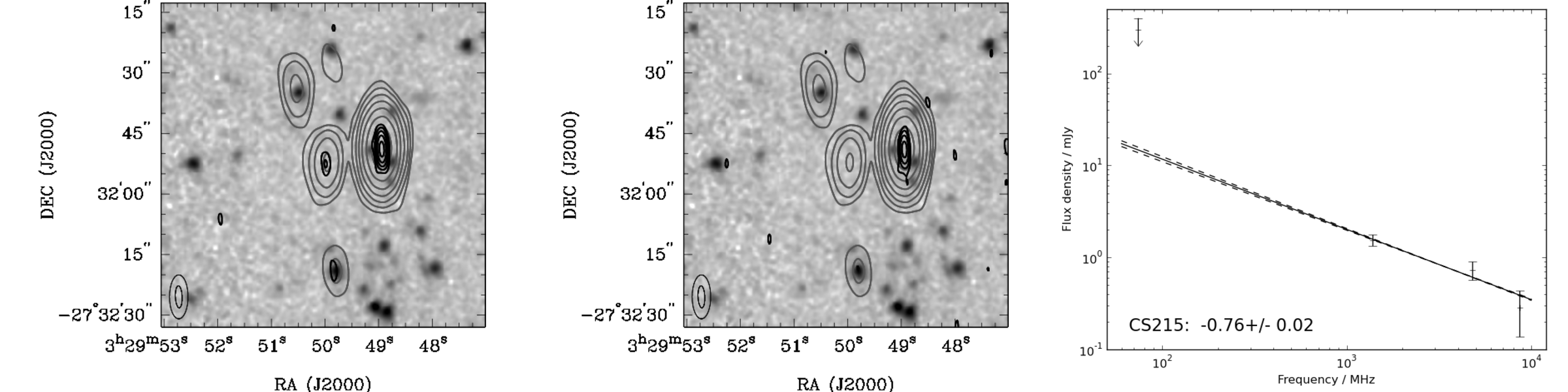}
\caption{(continued)}
\end{figure*}

\begin{figure*}
\includegraphics[width=\linewidth]{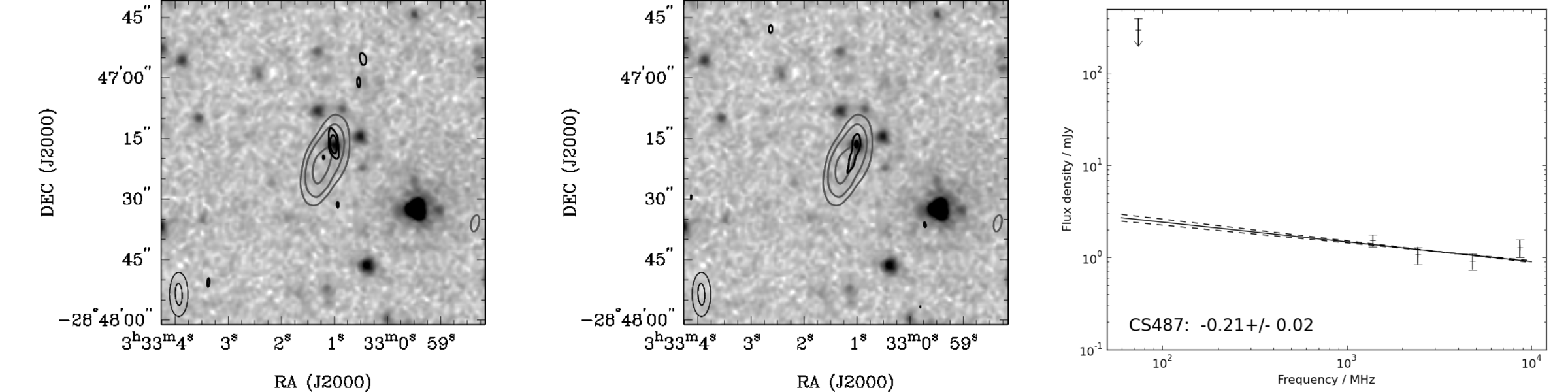}
\caption{The high-frequency images of CS487 clearly show an
  association of the radio emission with the infrared source
  SWIRE3\_J033300.99-284716.6, and hence is no longer classified as
  IFRS.}
\label{fig:cs487}
\end{figure*}

\end{document}